\documentclass[epj,nopacs]{svjour}
\usepackage{lineno}
\usepackage{amsmath}
\usepackage{amssymb}
\usepackage{epsfig}
\usepackage{color}

\newcommand{\beq}  {\begin{equation}}
\newcommand{\eeq}  {\end{equation}}

\usepackage{ifpdf}
\ifpdf
\DeclareGraphicsExtensions{.pdf, .jpg, .tif}
\usepackage[%
  pdftitle={Transverse-target-spin asymmetry in exclusive $\omega$-meson electroproduction },%
  pdfauthor={The HERMES Collaboration},%
  pdfsubject={HERMES omega AUT},%
  pdfstartview=FitH,%
  bookmarks=true,%
  bookmarksopen=true,%
  breaklinks=true,%
  colorlinks=true,%
  linkcolor=blue,anchorcolor=blue,%
  citecolor=blue,filecolor=blue,%
  menucolor=blue,pagecolor=blue,%
  urlcolor=blue]{hyperref}
\else
\DeclareGraphicsExtensions{.eps, .jpg}
\usepackage[%
  breaklinks=true,%
  colorlinks=true,%
  linkcolor=blue,anchorcolor=blue,%
  citecolor=blue,filecolor=blue,%
  menucolor=blue,pagecolor=blue,%
  urlcolor=blue]{hyperref}
\fi

\begin{document}
\hugehead
%\linenumbers

\title{
  Transverse-target-spin asymmetry in exclusive $\omega$-meson electroproduction\\}

\author{ 
The HERMES Collaboration \medskip \\
A.~Airapetian$^{15,18}$,
N.~Akopov$^{29}$,
Z.~Akopov$^{7}$,
E.C.~Aschenauer$^{8}$,
W.~Augustyniak$^{28}$,
A.~Avetissian$^{29}$,
S.~Belostotski$^{21}$,
H.P.~Blok$^{20,27}$,
A.~Borissov$^{7}$,
V.~Bryzgalov$^{22}$,
G.P.~Capitani$^{13}$,
G.~Ciullo$^{11,12}$,
M.~Contalbrigo$^{11}$,
P.F.~Dalpiaz$^{11,12}$,
W.~Deconinck$^{7}$,
R.~De~Leo$^{2}$,
E.~De~Sanctis$^{13}$,
M.~Diefenthaler$^{10,17}$,
P.~Di~Nezza$^{13}$,
M.~D\"uren$^{15}$,
G.~Elbakian$^{29}$,
F.~Ellinghaus$^{6}$,
L.~Felawka$^{25}$,
S.~Frullani$^{23,24}$,
D.~Gabbert$^{8}$,
G.~Gapienko$^{22}$,
V.~Gapienko$^{22}$,
V.~Gharibyan$^{29}$,
F.~Giordano$^{11,12,17}$,
S.~Gliske$^{18}$,
D.~Hasch$^{13}$,
M.~Hoek$^{16}$,
Y.~Holler$^{7}$,
A.~Ivanilov$^{22}$,
H.E.~Jackson$^{1}$,
S.~Joosten$^{14}$,
R.~Kaiser$^{16}$,
G.~Karyan$^{29}$,
T.~Keri$^{15}$,
E.~Kinney$^{6}$,
A.~Kisselev$^{21}$,
V.~Korotkov$^{22}$,
V.~Kozlov$^{19}$,
V.G.~Krivokhijine$^{9}$,
L.~Lagamba$^{2}$,
L.~Lapik\'as$^{20}$,
I.~Lehmann$^{16}$,
P.~Lenisa$^{11,12}$,
W.~Lorenzon$^{18}$,
B.-Q.~Ma$^{3}$,
S.I.~Manaenkov$^{21}$,
Y.~Mao$^{3}$,
B.~Marianski$^{28}$,
H.~Marukyan$^{29}$,
Y.~Miyachi$^{26}$,
A.~Movsisyan$^{11,29}$,
V.~Muccifora$^{13}$,
Y.~Naryshkin$^{21}$,
A.~Nass$^{10}$,
M.~Negodaev$^{8}$,
W.-D.~Nowak$^{8}$,
L.L.~Pappalardo$^{11,12}$,
R.~Perez-Benito$^{15}$,
A.~Petrosyan$^{29}$,
P.E.~Reimer$^{1}$,
A.R.~Reolon$^{13}$,
C.~Riedl$^{8,17}$,
K.~Rith$^{10}$,
G.~Rosner$^{16}$,
A.~Rostomyan$^{7}$,
J.~Rubin$^{17,18}$,
D.~Ryckbosch$^{14}$,
Y.~Salomatin$^{22}$,
G.~Schnell$^{4,5,14}$,
B.~Seitz$^{16}$,
T.-A.~Shibata$^{26}$,
M.~Statera$^{11,12}$,
E.~Steffens$^{10}$,
J.J.M.~Steijger$^{20}$,
F.~Stinzing$^{10,}$\footnote{deceased},
S.~Taroian$^{29}$,
A.~Terkulov$^{19}$,
R.~Truty$^{17}$,
A.~Trzcinski$^{28}$,
M.~Tytgat$^{14}$,
Y.~Van~Haarlem$^{14}$,
C.~Van~Hulse$^{4,14}$,
V.~Vikhrov$^{21}$,
I.~Vilardi$^{2}$,
C.~Vogel$^{10}$,
S.~Wang$^{3}$,
S.~Yaschenko$^{7,10}$,
S.~Yen$^{25}$,
B.~Zihlmann$^{7}$,
P.~Zupranski$^{28}$
}

\institute{ 
$^1$Physics Division, Argonne National Laboratory, Argonne, Illinois 60439-4843, USA\\
$^2$Istituto Nazionale di Fisica Nucleare, Sezione di Bari, 70124 Bari, Italy\\
$^3$School of Physics, Peking University, Beijing 100871, China\\
$^4$Department of Theoretical Physics, University of the Basque Country UPV/EHU, 48080 Bilbao, Spain\\
$^5$IKERBASQUE, Basque Foundation for Science, 48013 Bilbao, Spain\\
$^6$Nuclear Physics Laboratory, University of Colorado, Boulder, Colorado 80309-0390, USA\\
$^7$DESY, 22603 Hamburg, Germany\\
$^8$DESY, 15738 Zeuthen, Germany\\
$^9$Joint Institute for Nuclear Research, 141980 Dubna, Russia\\
$^{10}$Physikalisches Institut, Universit\"at Erlangen-N\"urnberg, 91058 Erlangen, Germany\\
$^{11}$Istituto Nazionale di Fisica Nucleare, Sezione di Ferrara, 44122 Ferrara, Italy\\
$^{12}$Dipartimento di Fisica e Scienze della Terra, Universit\`a di Ferrara, 44122 Ferrara, Italy\\
$^{13}$Istituto Nazionale di Fisica Nucleare, Laboratori Nazionali di Frascati, 00044 Frascati, Italy\\
$^{14}$Department of Physics and Astronomy, Ghent University, 9000 Gent, Belgium\\
$^{15}$II. Physikalisches Institut, Justus-Liebig Universit\"at Gie{\ss}en, 35392 Gie{\ss}en, Germany\\
$^{16}$SUPA, School of Physics and Astronomy, University of Glasgow, Glasgow G12 8QQ, United Kingdom\\
$^{17}$Department of Physics, University of Illinois, Urbana, Illinois 61801-3080, USA\\
$^{18}$Randall Laboratory of Physics, University of Michigan, Ann Arbor, Michigan 48109-1040, USA \\
$^{19}$Lebedev Physical Institute, 117924 Moscow, Russia\\
$^{20}$National Institute for Subatomic Physics (Nikhef), 1009 DB Amsterdam, The Netherlands\\
$^{21}$B.P. Konstantinov Petersburg Nuclear Physics Institute, Gatchina, 188300 Leningrad Region, Russia\\
$^{22}$Institute for High Energy Physics, Protvino, 142281 Moscow Region, Russia\\
$^{23}$Istituto Nazionale di Fisica Nucleare, Sezione di Roma, Gruppo Collegato Sanit\`a, 00161 Roma, Italy\\
$^{24}$Istituto Superiore di Sanit\`a, 00161 Roma, Italy\\
$^{25}$TRIUMF, Vancouver, British Columbia V6T 2A3, Canada\\
$^{26}$Department of Physics, Tokyo Institute of Technology, Tokyo 152, Japan\\
$^{27}$Department of Physics and Astronomy, VU University, 1081 HV Amsterdam, The Netherlands\\
$^{28}$National Centre for Nuclear Research, 00-689 Warsaw, Poland\\
$^{29}$Yerevan Physics Institute, 375036 Yerevan, Armenia\\
}

\date{DESY Report 15-149 / Compiled: \today /  Version: 5.0 (final)}

\authorrunning{The HERMES Collaboration}

\abstract{
Hard exclusive electroproduction of  $\omega$ 
mesons is studied with the HERMES spectrometer at the DESY laboratory  
 by scattering 27.6~GeV positron and electron beams off a transversely
polarized hydrogen target. The amplitudes of five  azimuthal modulations 
of the single-spin asymmetry of the cross section with respect  to the transverse proton polarization are measured.
They are determined in the entire kinematic region as well as for two bins in photon virtuality and momentum transfer to the nucleon. Also, a separation of asymmetry amplitudes into longitudinal and transverse components
is done. These results are compared to a phenomenological model that includes  the pion  pole
contribution. Within this model, the data favor  a positive $\pi\omega$ transition form
factor.} 
\maketitle

%%%%%%%%%%%%%%%%%%%%%%%%%%%%%%%%%%%%%%%%%
\subsection*{Introduction}
\label{intro}
In the framework of quantum chromodynamics (QCD), hard exclusive meson leptoproduction
on a longitudinally or transversely
polarized proton target provides important information about the spin structure of the nucleon. 
The process amplitude is a convolution of the lepton-quark
hard-scat\-te\-ring subprocess amplitude with soft  ha\-dronic matrix elements describing
the structure of the nucleon and that of the meson. 
Here, factorization is proven rigorously only if the lepton-quark interaction 
is mediated by a longitudinally polarized virtual photon~\cite{Collins:1996fb,Radyushkin:1996ru}.
The soft hadronic matrix elements  describing the nucleon contain generalized parton distributions (GPDs) 
to parametrize its partonic structure. Hard exclusive
production of vector mesons is described by GPDs $H^f$ and $E^f$,
where $f$ denotes a quark flavor or a gluon. 
These ``unpolarized'', i.e., parton-helicity-nonflip distributions
describe the photon-parton 
interaction with conservation and flip of nucleon helicity, respectively. 
Both are of special
interest, as they are related to the total angular momentum of
partons, $J^f$~\cite{Ji1}. The GPDs $H^f$ are well constrained
by existing experimental data. 
The GPDs $E^{u}$ and $E^{d}$ for up and down quarks, respectively, are partially constrained by nucleon form-factor data~\cite{Diehl_KrohlEPJC73}, 
while experimental information on sea-quark GPD $E^{\text{sea}}$ and gluon GPD $E^{g}$ is scarce. 
For a recent review on the status of GPD determinations, see Ref.~\cite{tr2014}. 
In contrast to leptoproduction of vector mesons with an unpolarized
target, which is mainly sensitive to GPDs $H^f$, 
vector-meson leptoproduction off a transversely polarized nucleon is sensitive to
the interference between two amplitudes containing $H^f$ and $E^f$,
respectively, and thus opens access to $E^f$.

For a transversely polarized virtual photon mediating the lepton-quark
interaction, there exists no rigorous proof of collinear factorization. 
In the QCD-inspired phenomenological 
``GK'' model~\cite{Goloskokov:2005sd,Goloskokov:2007nt,Goloskokov:2013mba} however,
factorization is also assumed for the transverse amplitudes. 
In this so-called modified perturbative approach~\cite{Botts:1989kf},
infrared singularities occurring in these amplitudes are regularized by
quark transverse momenta in the subprocess, while the partons are
still emitted and reabsorbed collinearly by the nucleon. By using the
quark transverse momenta in the subprocess, the transverse size of the
meson is effectively taken into account. Using this approach, the GK model describes
cross sections, spin density matrix elements (SDMEs), and spin asymmetries in
exclusive vector-meson production for values of Bjorken-$x$ below
0.2~\cite{Goloskokov:2005sd,Goloskokov:2007nt,Goloskokov:2013mba}. 
The GPDs parametrized in the GK model were used in calculations of deeply virtual Compton scattering (DVCS) amplitudes, 
which led to good agreement with most DVCS measurements over a wide kinematic range~\cite{dvcs}.
In the most recent version of the model, the $\gamma^* \pi \omega$ vertex function in the one-pion-exchange contribution 
is identified with the $\pi \omega$ transition form factor~\cite{piongk}. Its magnitude is
determined in a model-dependent way, while its unknown sign may be
determined from comparisons with experimental data on spin asymmetries in
hard exclusive leptoproduction. 

Measurements of hard-exclusive production of various types of 
mesons are complementary to DVCS,
as they allow access to various flavor combinations of
GPDs. Previous HERMES publications on measurements of azimuthal
transverse-target-spin asymmetries include results on exclusive
production of $\rho^0$~\cite{hermes1} and $\pi^+$
mesons~\cite{hermes2} as well as on DVCS \cite{hermes3}. 

In the present paper, the azimuthal modulations of the
transverse-target-spin asymmetry in the cross section of exclusive
electroproduction of $\omega$ mesons are studied. 
The available data allow for an estimation of 
the kinematic dependence of the measured asymmetry amplitudes 
on photon virtuality and  four-momentum
transfer to the nucleon. 
The measured asymmetry amplitudes are compared to the most recent calculations
of the GK model using either possible sign of the $\pi \omega$ transition
form factor.

\subsection*{Data collection and process identification}

The data were accumulated  with the HERMES forward
spectrometer~\cite{hermes5} during the running period 2002-2005.
The 27.6 GeV positron (electron) beam was scattered off a transversely
polarized hydrogen target, with the average magnitude $P_{T}$ of the
proton-polarization component $\vec{P_T}$ perpendicular to the beam direction 
being equal to $0.72$. The lepton beam was
longitudinally polarized, and in the analysis the data set is
beam-helicity balanced. The $\omega$ meson is produced in the  reaction
\begin{linenomath}
\beq
e + p \to e + p +  \omega,
\label{omprod}
\eeq
\end{linenomath}
with a branching ratio $Br = 89.1 \%$ for the $\omega$ decay
\begin{linenomath}
\beq
\omega \to \pi^+ + \pi^- + \pi^0,~\pi^0 \to2\gamma.
\label{omdecay}
\eeq
\end{linenomath}

\begin{figure}
\centering
\includegraphics[width=8cm]{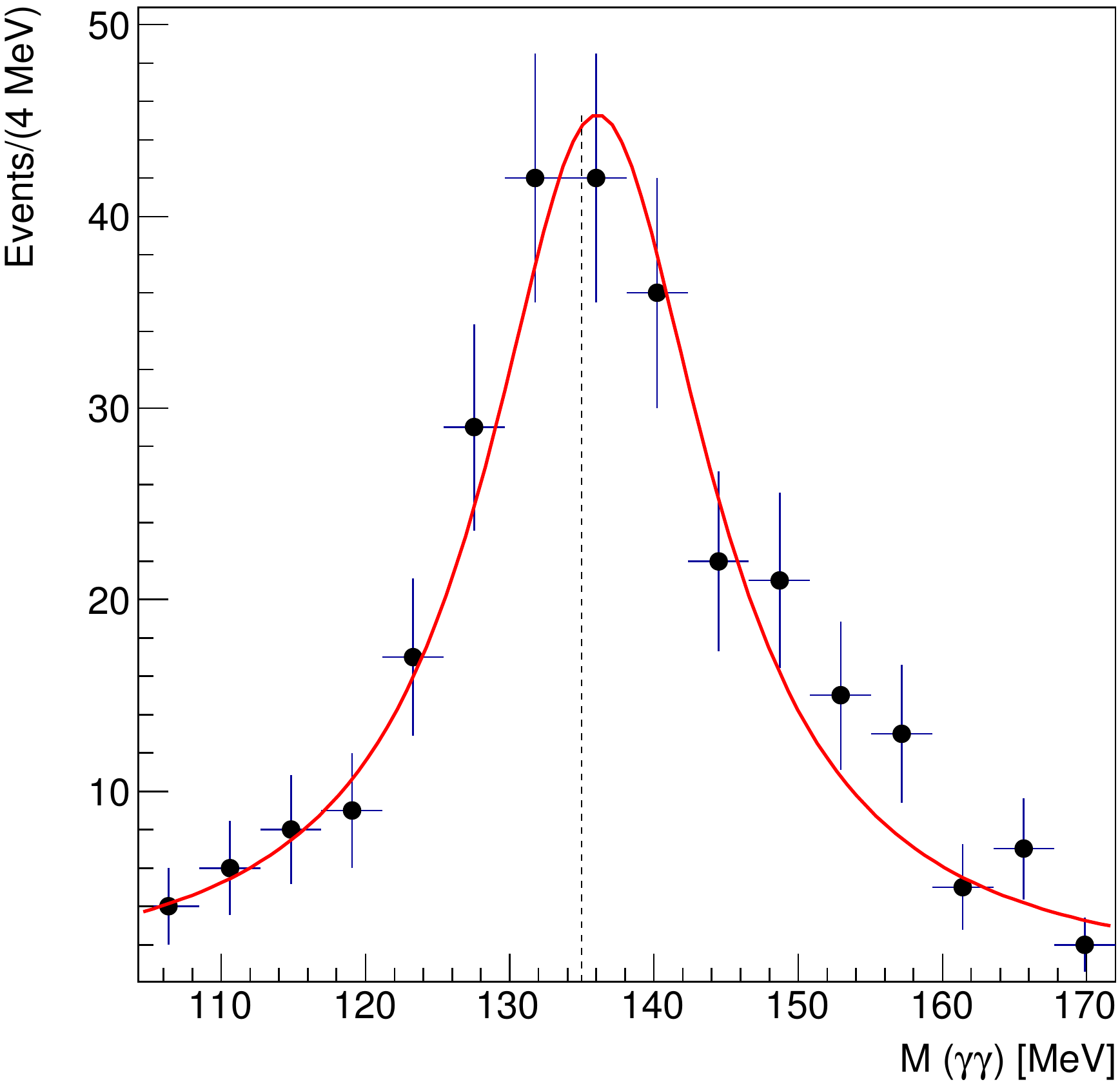}
\caption{Two-photon invariant-mass distribution after application of
all criteria to select exclusively produced $\omega$  mesons. 
The Breit--Wigner fit to the mass distribution is shown as a continuous  line
and the vertical dashed line indicates the PDG value of the $\pi^{0}$ mass~\cite{pdg}.}
\label{fig:invmassgamma}
\end{figure}

The same requirements to select exclusively produced $\omega$ mesons
as in Ref.~\cite{omega_sdme} are applied. The candidate events for
exclusive $\omega$-meson production are required to have exactly three
charged tracks, i.e., the scattered lepton and two oppositely charged
pions, and at least two clusters in the calorimeter not associated
with a charged track. The $\pi^0$ meson is reconstructed from 
two photon clusters with an invariant mass $M(\gamma \gamma)$ in the
interval $ 0.11$~GeV $< M(\gamma \gamma) <0.16 $~GeV. Its distribution
is shown in Fig.~\ref{fig:invmassgamma}, where the fit with a
Breit--Wigner function yields $136.1\pm0.8$~MeV ($19\pm2$~MeV) for the mass
(width). The charged hadrons and leptons are identified through the
combined responses of four particle-identification detectors~\cite{hermes5}. The three-pion invariant mass is calculated 
as $ M(\pi^{+}\pi^{-}\pi^{0}) = $ $\sqrt{(p_{\pi^+} + p_{\pi^-} +
  p_{\pi^0})^2}$, where $p_{\pi}$ are the four-momenta of the charged and 
neutral pions. Events containing  $\omega$ mesons are selected through
the requirement 0.71~GeV $< M(\pi^{+}\pi^{-}\pi^{0}) <$ 0.87~GeV. 

Further event-selection requirements are the following: 
\begin{itemize}
\item[(i)] 1.0~GeV$^2< Q^2 <$ 10.0~GeV$^2$, where $Q^2$
  represents the negative square of the virtual-photon
  four-momentum. The lower value is applied in order to facilitate the
  application of perturbative QCD, while the upper value delimits the
  measured phase space; 
\item[(ii)] $-t^{\prime} <$ 0.2~GeV$^2$ in order to improve
  exclusivity, where $t'=t - t_{min}$, $t$ is the squared 
  four-momentum transfer to the nucleon and $-t_{min}$ represents the
  smallest kinematically allowed value of $-t$ at fixed virtual-photon
  energy and $Q^{2}$; 
\item[(iii)] $W>$ 3~GeV in order to be outside of the resonance region
  and $W <$ 6.3~GeV in order to clearly delimit the kinematic phase
  space, where $W$ is the invariant mass of the photon-nucleon system; 
\item[(iv)] the scattered-lepton energy lies above 3.5~GeV in order to avoid a bias originating from the trigger. 
\end{itemize}

\begin{figure}
\centering
\includegraphics[width=8.5cm]{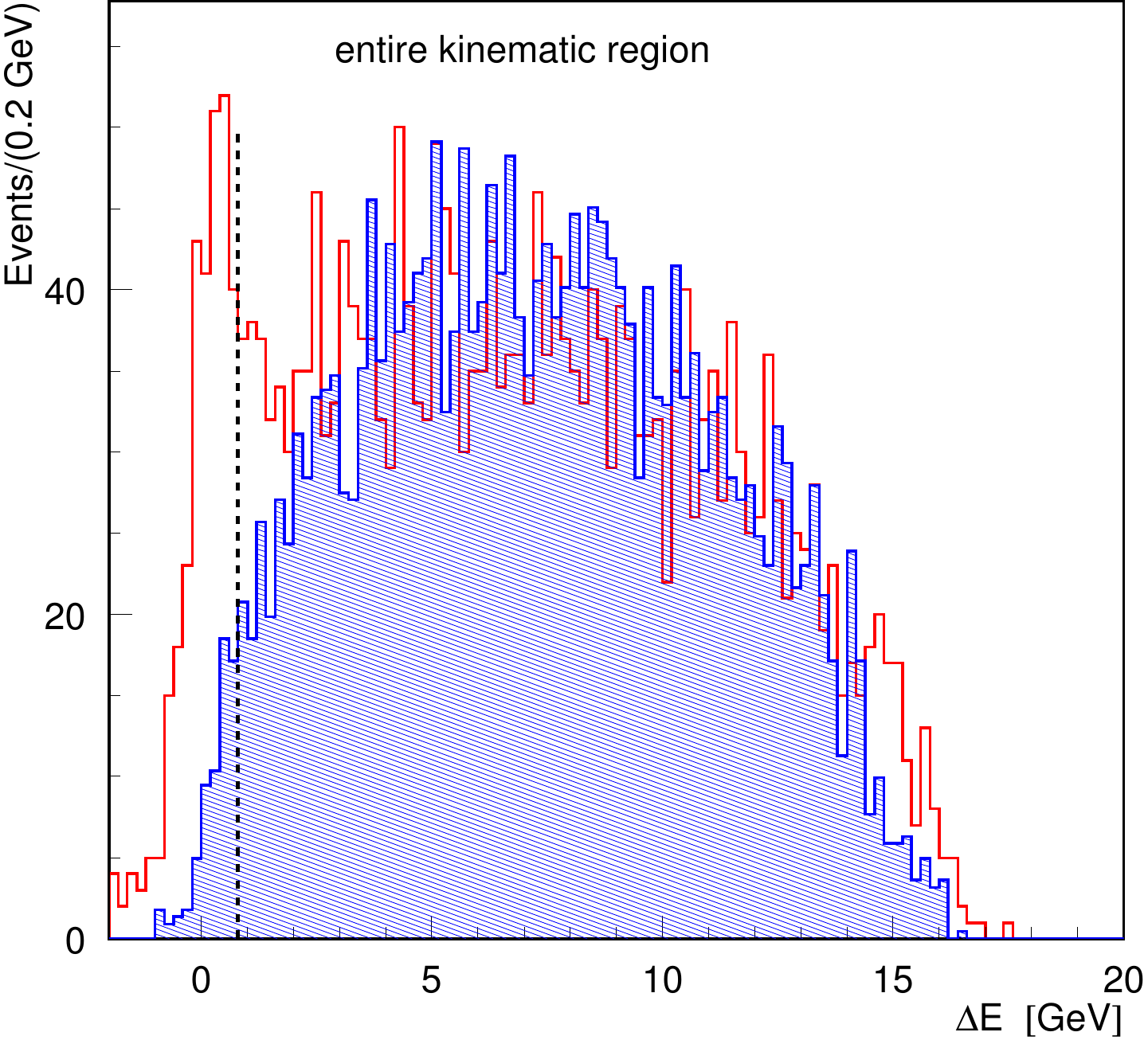}
\caption{Missing-energy distribution for exclusive $\omega$-meson production. 
The unshaded histogram shows experimental data, while the
shaded area shows the distribution obtained from a PYTHIA simulation of
the SIDIS background. The vertical dashed line denotes the upper limit of the exclusive region.}
\label{fig:deltaE_om}
\end{figure}

In order to isolate exclusive production, the energy not accounted for by
the leptons and the three pions must be zero within the experimental
resolution. We require the missing energy to be in the interval $-$1.0~GeV $<
\Delta E <$ 0.8~GeV, which is referred to as ``exclusive region" in the
following.
Here, the missing energy is calculated as 
$ \Delta E = \frac{M^{2}_{X} -M^{2}_{p}}{2 M_{p}}$, with $M_{p}$ being
the proton mass and
$M^{2}_{X}=(p+q-p_{\pi^+}-p_{\pi^-}-p_{\pi^0})^{2}$ the missing-mass
squared, where ${p}$ and ${q}$ are the four-momenta of target nucleon
and virtual photon, respectively. The distribution of the missing energy $\Delta E$ is shown in Fig.~\ref{fig:deltaE_om}. 
It exhibits a clearly visible exclusive peak centered about $\Delta E$ = 0. The shaded
area represents semi-inclusive deep-inelastic scattering (SIDIS) 
background events obtained from a PYTHIA~\cite{pythia} Monte-Carlo simulation that
is normalized to the data in the region 2~GeV $< \Delta E<$ 20~GeV. 
The simulation is  used to determine the fraction of background under the
exclusive peak. This fraction is calculated as the ratio of the number of background events to the
total number of events and amounts to  about $21\%$.

\begin{figure}
\centering
\includegraphics[width=8cm]{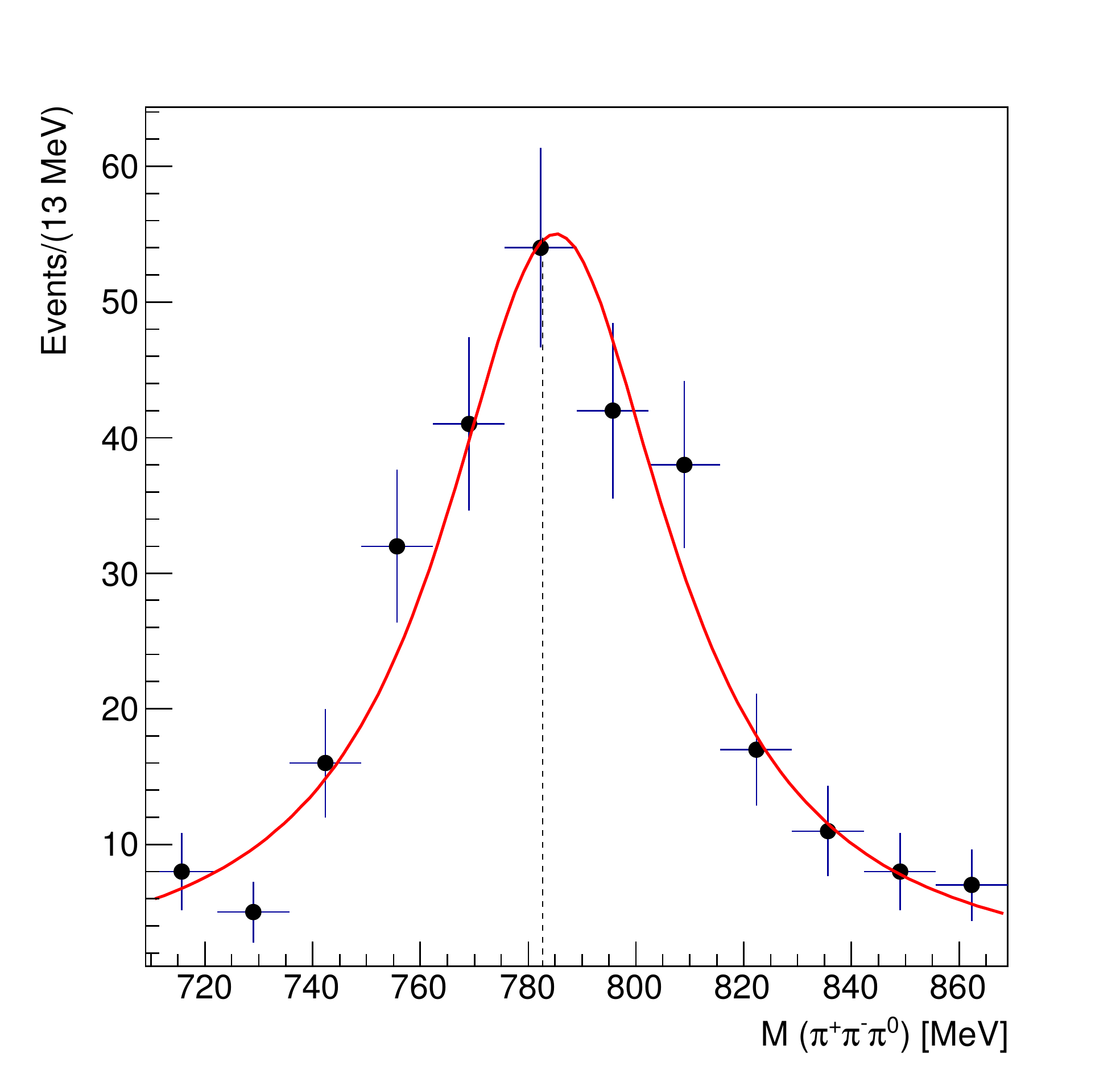}
\caption{The $ \pi^+ \pi^- \pi^0$ invariant-mass distribution after application of
all criteria to select exclusively produced $\omega$ mesons. 
The Breit--Wigner fit to the mass distribution is shown as a continuous  line
and the vertical dashed line indicates the PDG value of the $\omega$ mass~\cite{pdg}.} 
\label{fig:omega:invmassomega}
\end{figure}

After application of all these constraints, the sample contains 279 exclusively produced  $\omega$
mesons. This data sample is referred to in the following  as data in the ``entire kinematic region''.
The $\pi^{+}\pi^{-}\pi^{0}$ invariant-mass distribution for this data sample is shown in Fig.~\ref{fig:omega:invmassomega}.
A Breit--Wigner fit yields $785\pm2$~MeV ($52\pm5$~MeV) for the mass (width).

\subsection*{Extraction of the asymmetry amplitudes}
\begin{figure}
\centering
\includegraphics[width=.45\textwidth,angle=0]{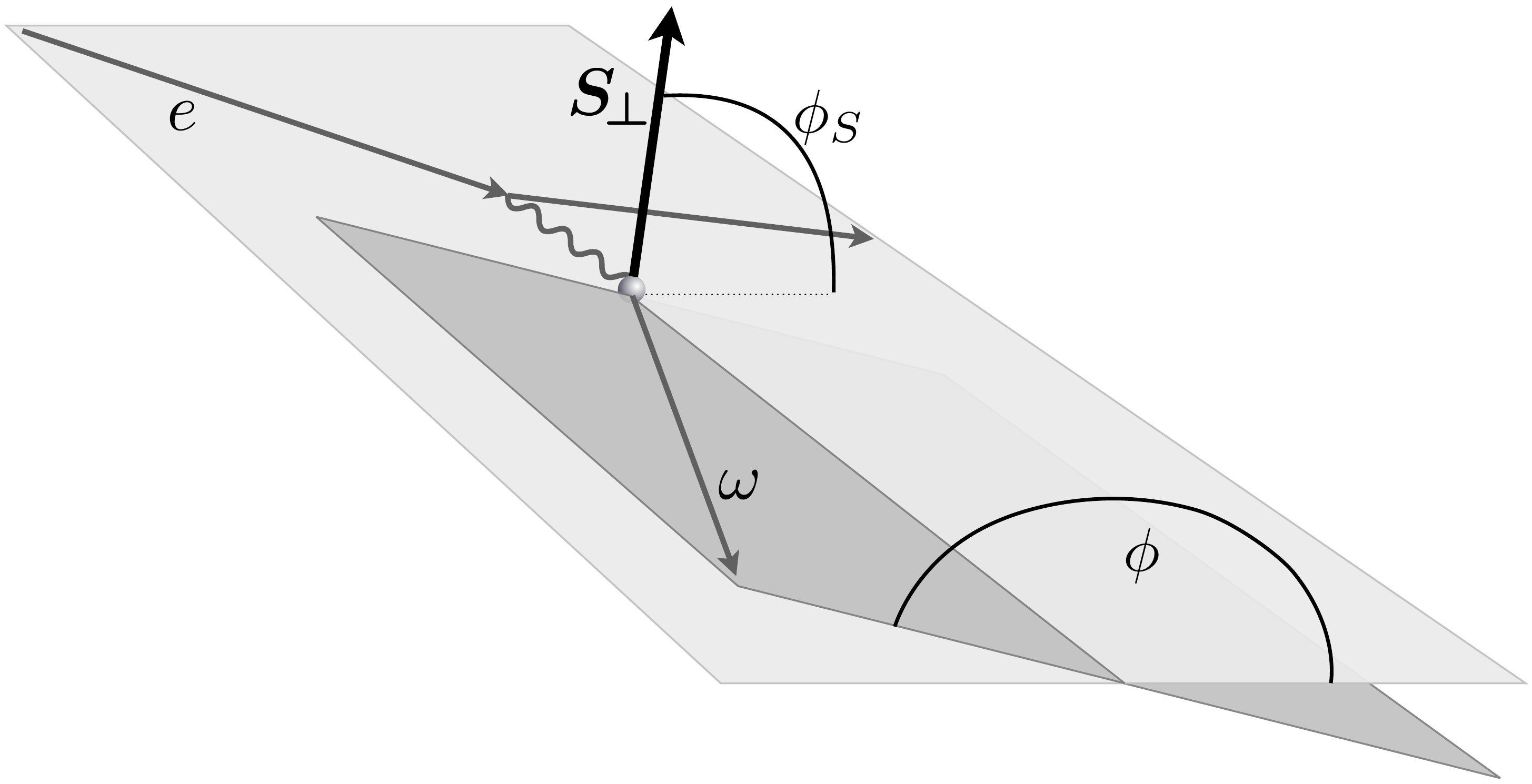}
\caption{Lepton-scattering and $\omega$-production planes together
  with the azimuthal angles $\phi$ and $\phi_S$.}
%  , where $\vec{P_h}$
%  denotes the three-momentum of the produced $\omega$ meson.}
\label{fig:angle_def}
\end{figure}

The cross section for hard exclusive leptoproduction of a vector meson on
a transversely polarized proton target, written in terms of 
polarized photo-absorption cross sections and interference terms, is given by Eq.~(34) in
Ref.~\cite{Diehl2}. 
In this equation, the transverse-target-spin asymmetry $A_{UT}$ is decomposed
into a Fourier series of terms involving $\sin(m\phi \pm \phi_{S})$, with $m=0,...,3$. 
The angles $\phi$ and $\phi_S$ are the azimuthal angles of the
$\omega$-production plane and of the component $\vec{S}_{\perp}$ of the 
transverse nucleon polarization vector that is orthogonal to the virtual-photon direction. They are measured around
the virtual--photon direction and with respect to the
lepton-scattering plane (see Fig.~\ref{fig:angle_def}). 
These definitions are in accordance with the Trento Conventions~\cite{Bacchetta1}.
For the HERMES kinematics and acceptance  in  exclusive  $\omega$  production,  
$\sin\theta_{\gamma^{*}}<$ 0.1 and $\cos\theta_{\gamma^{*}}>$ 0.99, which can be 
approximated by $\sin\theta_{\gamma^{*}} \approx$ 0 and $\cos\theta_{\gamma^{*}} \approx$ 1.
Here, $\theta_{\gamma^{*}}$ is the angle between the lepton-beam and
virtual-photon directions.

In this approximation, the angular-dependent part of Eq.~(34) in
Ref.~\cite{Diehl2} for an unpolarized beam reads: 
\begin{linenomath}
\beq
\begin{split}
{\cal W}(\phi,\phi_{S}) & = 1 +A^{\cos(\phi)}_{UU}\cos(\phi)+ A^{\cos(2\phi)}_{UU}\cos(2\phi)\\
& \quad +S_{\perp} [A^{\sin(\phi+\phi_{S})}_{UT}\sin(\phi+\phi_{S}) \\
& \qquad \; \, +A^{\sin(\phi-\phi_{S})}_{UT}\sin(\phi-\phi_{S}) \\ 
& \qquad \; \, +A^{\sin(\phi_{S})}_{UT}\sin(\phi_{S}) \\ 
& \qquad \; \, +A^{\sin(2\phi-\phi_{S})}_{UT}\sin(2\phi-\phi_{S}) \\
& \qquad \; \, +A^{\sin(3\phi-\phi_{S})}_{UT}\sin(3\phi-\phi_{S})], 
\label{eq_W}
\end{split}
\eeq
\end{linenomath}
where $S_{\perp}=|\vec{S}_{\perp}|$. Here, $A_{UU}$ and $A_{UT}$ denote the amplitudes of the corresponding
cosine and sine modulations as given in their superscripts. The first
letter in the subscript denotes unpolarized beam and the second letter
$U$ ($T$) denotes unpolarized (transversely polarized) target. The
above approximation in conjunction with the additional factor
$\epsilon/2$ $\approx 0.4$, where $\epsilon$ is the ratio of fluxes of
longitudinal and transverse virtual photons, allows one to neglect the
contribution of the $\sin(2\phi+\phi_S)$ modulation, appearing in Eq. (34) of Ref.~\cite{Diehl2}. This
approximation also makes the angular dependence of $S_{\perp}$
disappear (see Eq.~(8) of Ref.~\cite{Diehl2}), and
$S_{\perp}\simeq P_T$=$0.72$ is used in the following. 
Note that the modulation $\sin(\phi-\phi_{S})$ is the only one that appears at leading twist.

\begin{figure*}[t!]
\centering
\includegraphics[width=18cm]{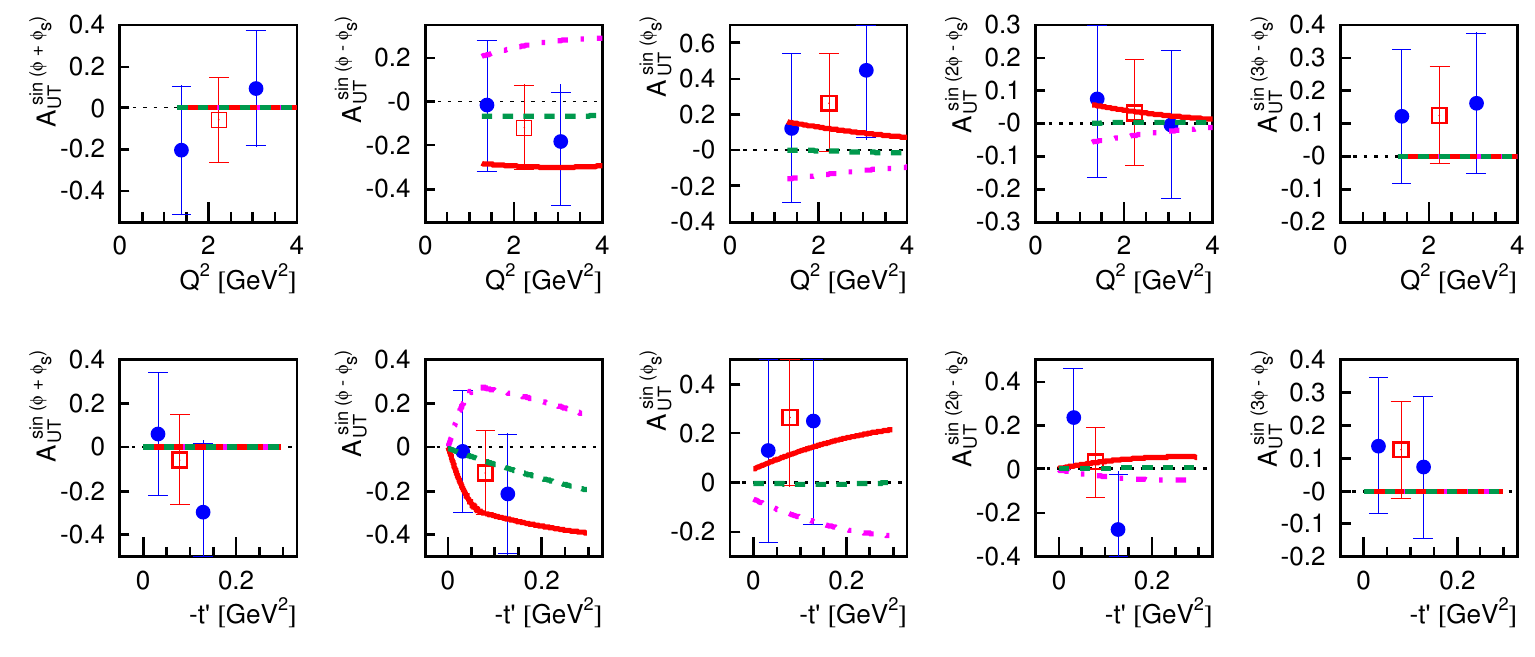}  
\caption{The five amplitudes describing the strength of the sine modulations of
the cross section for hard exclusive $\omega$-meson production. The  full circles show the data in two bins of $Q^2$ or $-t'$. The  open squares represent the results obtained for 
the entire kinematic region. 
The inner error bars represent the statistical uncertainties, while the outer ones indicate the statistical and systematic uncertainties added in quadrature. The results receive an additional 8.2\% scale uncertainty corresponding to the target-polarization uncertainty. 
The solid (dash-dotted) lines show the calculation of the  GK model~\cite{piongk,piongk1}  
for a positive (negative) $\pi\omega $ transition form factor, and the dashed lines are the model results without the pion pole.} 
\label{fig:omegakin:results}
\end{figure*}

For exclusive production of $\omega$ mesons decaying into three pions, the angular distribution of the latter can be 
decomposed into parts corresponding to longitudinally (L) and transversely (T) polarized
$\omega$ mesons: 
\begin{linenomath}
\beq
\begin{split} 
{\cal W}(\phi,\phi_S,\theta)& = \frac{3}{2}\, r^{04}_{00} \, \cos^2(\theta) \, w_L(\phi,\phi_S)\\
& \quad +  \frac{3}{4}\, (1-r^{04}_{00}) \, \sin^2(\theta) \, w_T(\phi,\phi_S).
\label{wlt1}
\end{split}
\eeq
\end{linenomath}
Here, $\theta$ is the polar angle of the unit vector normal to the $\omega$ decay plane in the $\omega$-meson rest frame, 
with the $z$-axis aligned opposite to the outgoing nucleon momentum~\cite{omega_sdme}.  
The pre-factors $r^{04}_{00}$ and ($1 - r^{04}_{00}$) represent  the fractional contribution to
the full cross section by longitudinally and transversely polarized $\omega$
mesons, respectively~\cite{omega_sdme}. 
The first (second) term on the right-hand side of Eq.~(\ref{wlt1})
represents the angular distribution of the longitudinally (transversely)
polarized $\omega$ mesons, with 
\begin{linenomath}
\begin{eqnarray}
\begin{split}
w_{L}(\phi,\phi_S)& = 1 + A_{UU,L}(\phi) + S_{\perp} A_{UT,L}(\phi,\phi_S),\\
\label{wl}
w_{T}(\phi,\phi_S)& = 1 + A_{UU,T}(\phi) + S_{\perp} A_{UT,T}(\phi,\phi_S). 
\label{wt}
\end{split}
\end{eqnarray}
\end{linenomath}
The functions $A_{UU,K}(\phi)$ and $A_{UT,K}(\phi,\phi_S)$, with $K$=$L$ and $K$=$T$ denoting longitudinal-separated and 
transverse-separated contributions, respectively, are decomposed into a Fourier series in complete analogy 
to Eq.~(\ref{eq_W}).  

The function ${\cal W}(\phi,\phi_{S})$ is fitted to the  experimental angular
distribution using an unbinned maximum likelihood method. Here and in
the following, the angle $\theta$ has to be added to the argument list
of the function $\cal{W}$, when applicable. The function to be
minimized is the negative of the logarithm of the likelihood function:

\begin{linenomath}
\beq
-\ln L({\cal R})=-\sum_{i=1}^{N}\ln\frac{{\cal W}({\cal R};\phi^{i},\phi_{S}^{i})}{\widetilde{\mathcal N}({\cal R})}.
\label{eq_log}
\eeq
\end{linenomath}
Here, $\cal R$ denotes the set of 7 asymmetry amplitudes of the unseparated fit 
or 14 asymmetry amplitudes of the longitudinal-to-transverse separated
fit and the sum runs over the $N$ experimental-data events. The normalization factor 
\begin{linenomath}
\beq
 \widetilde {\mathcal N}({\cal R})=\sum_{j=1}^{N_{MC}}{\cal W}({\cal R};\phi^{j},\phi_{S}^{j})   
\label{eq_norm} 
\eeq
\end{linenomath}
is determined using $N_{MC}$ events from a PYTHIA Monte-Carlo
simulation, which are generated according to an isotropic 
angular distribution and processed in the same way
as experimental data. 
The number of Monte-Carlo events in the exclusive region amounts to about 40,000.

Each asymmetry amplitude is corrected for the background asymmetry according to 
\begin{linenomath}
\beq
 A_{corr}=\frac{A_{meas}-f_{bg}A_{bg}}{1-f_{bg}}, 
\eeq
\end{linenomath}
where $A_{corr}$ is the corrected asymmetry amplitude, $A_{meas}$ is the
measured asymmetry amplitude, $f_{bg}$ is the fraction of the SIDIS
background and $A_{bg}$ is its asymmetry amplitude. While $A_{meas}$ is 
evaluated in the exclusive region, $A_{bg}$ is obtained
by extracting the asymmetry from the experimental SIDIS background in
the region 2~GeV $< \Delta  E <$ 20~GeV. 

The systematic uncertainty is obtained by adding in quadrature
two components. The first one, \(\Delta A_{corr}  = A_{corr}-A_{meas}\), 
is due to the correction by background amplitudes. In the most
conservative approach adopted here, it is estimated as the difference
between the asymmetry amplitudes $A_{corr}$ and $A_{meas}$. This
approach also covers the small uncertainty on $f_{bg}$. 
The second component accounts for effects from detector acceptance, efficiency, smearing, and misalignment. 
It is determined as described in Ref.~\cite{omega_sdme}. 
An additional scale uncertainty arises because of the systematic uncertainty on the target polarization, which
amounts to 8.2\%.

\subsection*{Results}

\begin{table}[b] 
\renewcommand{\arraystretch}{1.2}
\caption{\label{tab1} The amplitudes of the five sine and two cosine
modulations as determined in the entire kinematic region. 
The first uncertainty is statistical, the second  systematic. 
The results receive an additional 8.2\% scale uncertainty corresponding to the target-polarization uncertainty. } 
\centering
\begin{tabular}{|l|r|}
\hline 
 amplitude &     \\ 
\hline
$A^{\sin(\phi+\phi_{S})}_{UT}$  & $-$0.06 $\pm$ 0.20 $\pm$ 0.02 \\
$A^{\sin(\phi-\phi_{S})}_{UT}$  & $-$0.12 $\pm$ 0.19 $\pm$ 0.03 \\
$A^{\sin(\phi_{S})}_{UT}$       &  0.26 $\pm$ 0.27 $\pm$ 0.05 \\
$A^{\sin(2\phi-\phi_{S})}_{UT}$ &  0.03 $\pm$ 0.16 $\pm$ 0.01 \\
$A^{\sin(3\phi-\phi_{S})}_{UT}$ &  0.13 $\pm$ 0.15 $\pm$ 0.03 \\
$A^{\cos(\phi)}_{UU}$           & $-$0.01 $\pm$ 0.11 $\pm$ 0.10 \\
$A^{\cos(2\phi)}_{UU}$          & $-$0.17 $\pm$ 0.11 $\pm$ 0.05 \\
\hline
\end{tabular}
\end{table}

\begin{table*} 
\renewcommand{\arraystretch}{1.2}
\caption{\label{tab2} The definition of intervals and the mean values of the kinematic variables.}
\centering
\begin{tabular}{|c|c|c|c|c|}
\hline 
 bin  & $\langle Q^{2} \rangle$ [GeV$^2$] & $\langle-t'\rangle$ [GeV$^2$] &$\langle  W \rangle$ [GeV] & $\langle x_{B}\rangle$  \\
\hline 
\hline
  entire kinematic bin  & 2.24  &  0.079 & 4.80 & 0.092 \\
\hline
$1.00$ GeV$^2$$ <Q^{2}< 1.85$ GeV$^2$ & 1.39 &0.084 &4.69 & 0.064 \\
$1.85$ GeV$^2$$ <Q^{2}< 10.00$ GeV$^2$ &  3.07 &0.075 &4.91 & 0.012 \\ 
\hline
$0.00$ GeV$^2$$ <-t'< 0.07$ GeV$^2$ & 2.36  & 0.035 & 4.79 & 0.095 \\
$0.07$ GeV$^2$$ <-t'< 0.20$ GeV$^2$ & 2.11  & 0.128 & 4.81 & 0.088 \\
\hline
\end{tabular}
\end{table*}

\begin{table*} 
\renewcommand{\arraystretch}{1.2}
\centering
\caption{\label{tab3} Results on the kinematic dependences of the five asymmetry amplitudes $A_{UT}$ and two amplitudes $A_{UU}$. 
The first two columns correspond to the $-t'$ intervals $0.00 - 0.07 - 0.20$ GeV$^2$ and the last two columns 
to the $ Q^{2}$ intervals $1.00 - 1.85 - 10.00$~GeV$^2$. 
The first uncertainty is statistical, the second  systematic.
The results receive an additional 8.2\% scale uncertainty corresponding to the target-polarization uncertainty. } 
\begin{tabular}{|l|r|r|r|r|}
\hline 
 amplitude & $\langle-t'\rangle$ = 0.035 GeV$^{2}$ & $\langle-t'\rangle$ =
 0.128 GeV$^{2}$ & $\langle Q^{2}\rangle$ = 1.39 GeV$^{2}$ &$\langle Q^{2}\rangle$ =  3.07 GeV$^{2}$  \\ 
\hline
$A^{\sin(\phi+\phi_{S})}_{UT}$  &  0.06 $\pm$ 0.28 $\pm$ 0.04 & $-$0.30 $\pm$ 0.32 $\pm$ 0.10 & $-$0.21 $\pm$ 0.31 $\pm$ 0.05 &  0.10 $\pm$ 0.28 $\pm$ 0.03 \\
$A^{\sin(\phi-\phi_{S})}_{UT}$  & $-$0.02 $\pm$ 0.28 $\pm$ 0.03 & $-$0.22 $\pm$ 0.27 $\pm$ 0.06 & $-$0.02 $\pm$ 0.30 $\pm$ 0.03 & $-$0.18 $\pm$ 0.26 $\pm$ 0.03 \\
$A^{\sin(\phi_{S})}_{UT}$       &  0.13 $\pm$ 0.37 $\pm$ 0.03 &  0.25 $\pm$ 0.42 $\pm$ 0.05 &  0.12 $\pm$ 0.42 $\pm$ 0.02 &  0.45 $\pm$ 0.37 $\pm$ 0.12 \\
$A^{\sin(2\phi-\phi_{S})}_{UT}$ &  0.24 $\pm$ 0.22 $\pm$ 0.03 & $-$0.28 $\pm$ 0.26 $\pm$ 0.07 &  0.07 $\pm$ 0.24 $\pm$ 0.02 & $-$0.01 $\pm$ 0.23 $\pm$ 0.01 \\
$A^{\sin(3\phi-\phi_{S})}_{UT}$ &  0.14 $\pm$ 0.21 $\pm$ 0.01 &  0.07 $\pm$ 0.22 $\pm$ 0.03 &  0.12 $\pm$ 0.20 $\pm$ 0.04 &  0.16 $\pm$ 0.21 $\pm$ 0.02 \\
$A^{\cos(\phi)}_{UU}$           &  0.05 $\pm$ 0.15 $\pm$ 0.06 & $-$0.09 $\pm$ 0.17 $\pm$ 0.16 & $-$0.04 $\pm$ 0.15 $\pm$ 0.10 & $-$0.04 $\pm$ 0.16 $\pm$ 0.11 \\
$A^{\cos(2\phi)}_{UU}$          & $-$0.19 $\pm$ 0.15 $\pm$ 0.07 & $-$0.14 $\pm$ 0.17 $\pm$ 0.07 & $-$0.04 $\pm$ 0.15 $\pm$ 0.03 & $-$0.35 $\pm$ 0.17 $\pm$ 0.11  \\
\hline
\end{tabular}
\end{table*}

The results for the five $A_{UT}$ and two $A_{UU}$ amplitudes, as
determined in the entire kinematic region, are shown in
Table~\ref{tab1}. These results are presented in Table~\ref{tab3} in
two intervals of $Q^2$ and $-t'$, with the definition of intervals
together with the average values of the respective kinematic variables
given in Table~\ref{tab2}. The results for the five $A_{UT}$
amplitudes are also shown in Fig.~\ref{fig:omegakin:results}, in two
rows of five panels each, where the upper and lower rows show the
$Q^2$ and $-t'$ dependences, respectively. Each panel shows as two
filled circles the results in two kinematic bins, and as one open
square the result in the entire kinematic region. The results are
compared to calculations of the GK model~\cite{piongk,piongk1}, 
for both signs of the $\pi\omega$ form factor. For completeness, also  
the model prediction without the pion-pole contribution is included. 

The model predictions differ substantially upon sign change of the
$\pi\omega$ form factor for the two amplitudes
$A^{\sin(\phi-\phi_{S})}_{UT}$ and $A^{\sin(\phi_{S})}_{UT}$, in
particular when considering the $-t'$ dependence. 
The data seem to favor a positive $\pi\omega$ transition form factor. 

Asymmetry amplitudes can be written in terms of SDMEs, as shown in the appendix.
By using Eqs.~(\ref{a1_o}, \ref{a2_o}) and the earlier HERMES results on $\omega$ SDMEs~\cite{omega_sdme}, 
\begin{align}
\nonumber
A_{UU}^{\cos(\phi)}&=-0.13\pm0.04\pm0.08\\ 
\nonumber
A_{UU}^{\cos(2\phi)}&=-0.03\pm0.04\pm0.01
\end{align}
are obtained, which are consistent within uncertainties with the results shown in Table~\ref{tab1}.

\begin{figure*}
\centering   
\includegraphics[width=18cm]{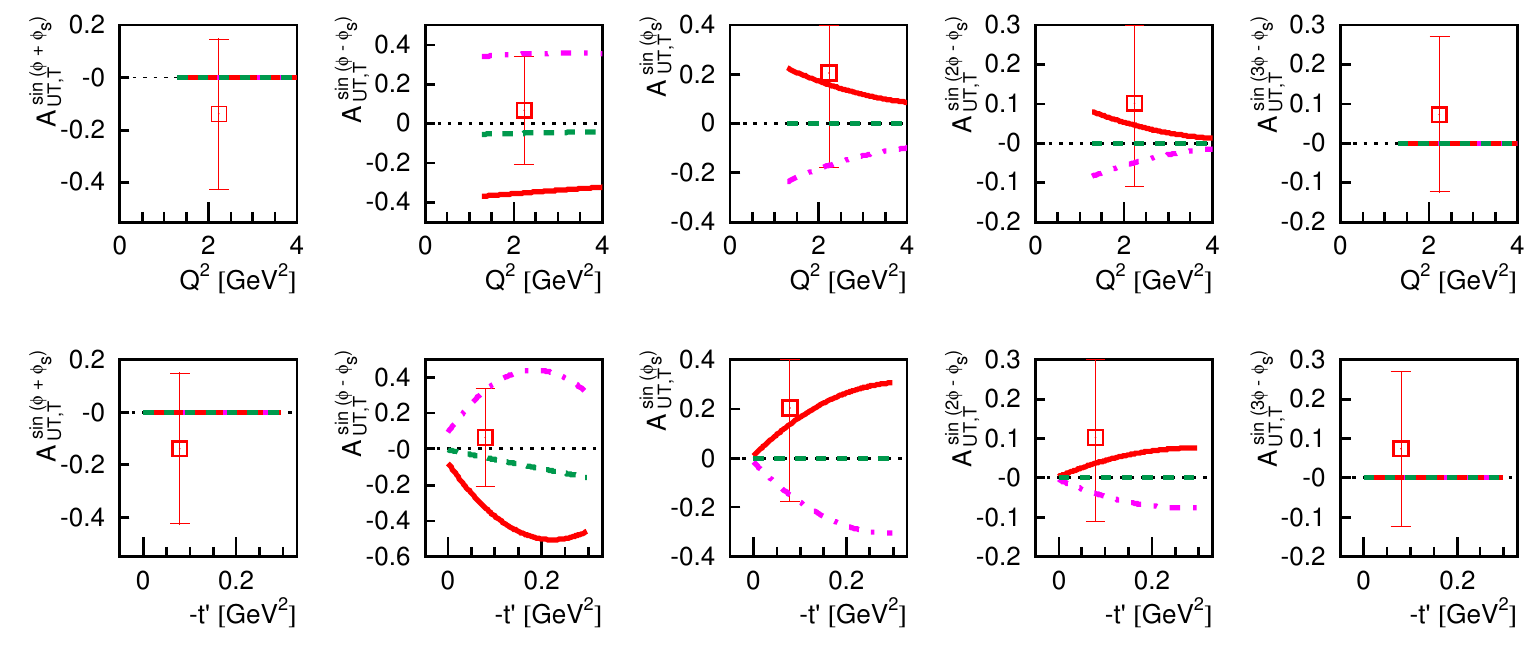}  
\caption{As Fig.~\ref{fig:omegakin:results}, but only for transversely polarized $\omega$ mesons.}
\label{autbaclt}
\end{figure*}

The cross section for exclusive production of transversely polarized
$\omega$ mesons dominates that for longitudinally polarized
ones~\cite{omega_sdme}. 
This is the reason why the 14-parameter fit used here leads to still acceptable
uncertainties for the results in the entire kinematic region on the transverse-separated
asymmetry amplitudes, while those for the longitudinal-separated
ones are so large that any interpretation is precluded. Also,
kinematic dependences can no longer be studied due to the large
uncertainties. Therefore, for the transverse-separated asymmetry
amplitudes only the results in the entire kinematic region  are shown in
Fig.~\ref{autbaclt} and Table~\ref{tab4} together
with the corresponding predictions of the GK model~\cite{piongk,piongk1}. Here, the 
large uncertainties prevent any conclusion on the sign of the $\pi \omega$
transition form factor.

\begin{table}[hbtc!] 
\renewcommand{\arraystretch}{1.2}
%\begin{center}
\caption{\label{tab4} Results on the five asymmetry amplitudes $A_{UT}$ and two amplitudes $A_{UU}$ in the entire kinematic region, 
but separated into longitudinal and transverse parts. The first column ($K=L$) gives the results for the longitudinal components, while 
the second column, ($K=T$), shows the results for the transverse components. 
The first uncertainty is statistical, the second  systematic.
The results receive an additional 8.2\% scale uncertainty corresponding to the target-polarization uncertainty. } 
\centering
\begin{tabular}{|l|r|r|}
\hline 
 amplitude & longitudinal ($K=L$)  &  transverse ($K=T$)   \\ 
\hline
$A^{\sin(\phi+\phi_{S})}_{UT,K}$  & $-$0.16 $\pm$ 0.92 $\pm$ 0.02 & $-$0.14 $\pm$ 0.29 $\pm$ 0.05 \\ 
$A^{\sin(\phi-\phi_{S})}_{UT,K}$  & $-$0.60 $\pm$ 0.81 $\pm$ 0.16 &  0.07 $\pm$ 0.27 $\pm$ 0.04 \\
$A^{\sin(\phi_{S})}_{UT,K}$       & $-$0.08 $\pm$ 1.06 $\pm$ 0.03 &  0.21 $\pm$ 0.38 $\pm$ 0.01 \\
$A^{\sin(2\phi-\phi_{S})}_{UT,K}$ & $-$0.38 $\pm$ 0.71 $\pm$ 0.11 &  0.10 $\pm$ 0.21 $\pm$ 0.02 \\
$A^{\sin(3\phi-\phi_{S})}_{UT,K}$ &  0.21 $\pm$ 0.56 $\pm$ 0.10 &  0.07 $\pm$ 0.20 $\pm$ 0.01 \\
$A^{\cos(\phi)}_{UU,K}$           &  0.53 $\pm$ 0.40 $\pm$ 0.08 & $-$0.16 $\pm$ 0.15 $\pm$ 0.12 \\
$A^{\cos(2\phi)}_{UU,K}$          &  0.60 $\pm$ 0.39 $\pm$ 0.17 & $-$0.37 $\pm$ 0.15 $\pm$ 0.10 \\
\hline
\end{tabular}
%\end{center} 
\end{table}

\subsection*{Summary}
In this Paper, results are reported on exclusive $\omega$ electroproduction off transversely
polarized protons in the kinematic region 1~GeV$^2 < Q^{2}< 10$ GeV$^2$ 
and 0.0 GeV$^2 < -t' < 0.2$ GeV$^2$. The amplitudes of seven azimuthal
modulations of the cross section for unpolarized beam are determined,
i.e., of two cosine modulations for unpolarized target and five  
sine modulations for transversely polarized target. Results are  
presented for the entire kinematic region as well as alternatively in
two bins of $-t'$ or  $Q^{2}$. 
Additionally, a separation into asymmetry amplitudes for the
production of longitudinally and transversely polarized $\omega$
mesons is done. 
A comparison of extracted asymmetry amplitudes to recent calculations of the
phenomenological model of Goloskokov and Kroll  favors a positive
sign of the $\pi \omega$ form factor.

\begin{acknowledgement}
{\bf Acknowledgments~}
We are grateful to Sergey Go\-los\-ko\-kov and Peter Kroll for fruitful
discussions on the comparison 
between our data and their model calculations.\\
We gratefully acknowledge the DESY management for its support and the staff
at DESY and the collaborating institutions for their significant effort.
This work was supported by 
the Ministry of Education and Science of Armenia;
the FWO-Flanders and IWT, Belgium;
the Natural Sciences and Engineering Research Council of Canada;
the National Natural Science Foundation of China;
the Alexander von Humboldt Stiftung,
the German Bundesministerium f\"ur Bildung und Forschung (BMBF), and
the Deutsche Forschungsgemeinschaft (DFG);
the Italian Istituto Nazionale di Fisica Nucleare (INFN);
the MEXT, JSPS, and G-COE of Japan;
the Dutch Foundation for Fundamenteel Onderzoek der Materie (FOM);
the Russian Academy of Science and the Russian Federal Agency for 
Science and Innovations;
the Basque Foundation for Science (IKERBASQUE) and the UPV/EHU under program UFI 11/55;
the U.K.~Engineering and Physical Sciences Research Council, 
the Science and Technology Facilities Council,
and the Scottish Universities Physics Alliance;
as well as the U.S.~Department of Energy (DOE) and the National Science Foundation (NSF).
\end{acknowledgement}

\begin{appendix}

\subsection*{Appendix: Relations between azimuthal asymmetry amplitudes and spin-density matrix elements}

The full information on vector-meson leptoproduction is contained in the differential cross section $\frac{d^3 \sigma}{dQ^2dtdx}$ 
and the SDMEs in the Diehl representation \cite{Diehl4}. 
Therefore, the azimuthal asymmetry amplitudes
can be expressed in terms of the SDMEs. For scattering off an unpolarized target, 
the asymmetry amplitudes can be written in 
terms of the Diehl 
SDMEs $u^{\mu_1\mu_2}_{\lambda_1 \lambda_2}$ or alternatively in terms of the Schilling--Wolf SDMEs $r^{n}_{ij}$ \cite{Schill} as
\begin{align}
\nonumber
A_{UU}^{\cos{\phi}}&=-2\sqrt{\epsilon(1+\epsilon)}\,\rm{Re}[u_{0+}]\\
&=\sqrt{2\epsilon(1+\epsilon)}\,[2r^5_{11}+r^5_{00}],
\label{a1_o}\\
\nonumber
A_{UU}^{\cos{2\phi}}&=-\epsilon \,\rm{Re}[u_{-+}]\\
&=-\epsilon \, [2r^1_{11}+r^1_{00}]. 
\label{a2_o}
\end{align}  
Here, the abbreviated notation 
\beq
u_{\lambda_1 \lambda_2}=u_{\lambda_1 \lambda_2}^{++}+u_{\lambda_1 \lambda_2}^{--}+ u_{\lambda_1 \lambda_2}^{00} 
\label{sum-u}   
\eeq
is used, where $\lambda_1$, $\lambda_2$ denote the virtual-photon helicities and $\mu_1$, $\mu_2$ the vector-meson helicities.
The symbol $\pm$ describes the virtual-photon or vector-meson helicities $\pm 1$, while the symbol $0$ describes longitudinal polarization.
Equations~(\ref{a1_o}, \ref{a2_o}) show that the asymmetry amplitudes can be
calculated from the 
Schilling--Wolf SDMEs obtained in Ref.~\cite{omega_sdme}.

For scattering off a transversely polarized target, the asymmetry amplitudes can be expressed in 
terms of the Diehl SDMEs $n^{\mu_1\mu_2}_{\lambda_1 \lambda_2}$ and $s^{\mu_1\mu_2}_{\lambda_1 \lambda_2}$ as 
\begin{align}
A_{UT}^{\sin(\phi+\phi_S)} & =(\epsilon/2)\,\rm{Im}[n_{-+}-s_{-+}],\label{def-a3}\\
A_{UT}^{\sin(\phi-\phi_S)}& =\rm{Im}[n_{++}+\epsilon n_{00}],\label{def-a4}\\  
A_{UT}^{\sin(\phi_S)}& =\sqrt{\epsilon(1+\epsilon)}\,\rm{Im}[n_{0+}-s_{0+}],\label{def-a5}\\
A_{UT}^{\sin(2\phi-\phi_S)}& =-\sqrt{\epsilon(1+\epsilon)}\,\rm{Im}[n_{0+}+s_{0+}],\label{def-a6}\\
A_{UT}^{\sin(3\phi-\phi_S)}& =-(\epsilon/2)\,\rm{Im}[n_{-+}+s_{-+}].\label{def-a7}
\end{align}
The abbreviated notations
\begin{align}
n_{\lambda_1 \lambda_2}&=n_{\lambda_1 \lambda_2}^{++}+n_{\lambda_1 \lambda_2}^{--}+ n_{\lambda_1 \lambda_2}^{00},\label{sum-n} \\  
s_{\lambda_1 \lambda_2}&=s_{\lambda_1 \lambda_2}^{++}+s_{\lambda_1 \lambda_2}^{--}+ s_{\lambda_1 \lambda_2}^{00} \label{sum-s} 
\end{align}
are analogous to those in Eq.~$(\ref{sum-u})$.
In this case, Schilling--Wolf SDMEs $r^{n}_{ij}$ \cite{Schill} are not defined. 

In order to get from Eqs.~(\ref{a1_o}, \ref{a2_o}) and  ~(\ref{def-a3}--\ref{def-a7}) expressions for the asymmetry amplitudes for the production of 
longitudinally polarized vector mesons, the terms with $\mu_1=\mu_2=0$ have to be retained in
Eqs.~(\ref{a1_o}--\ref{sum-s}),  
and the result has to be divided by the Schilling--Wolf SDME $r^{04}_{00}$. 
For instance, $A_{UT}^{\sin(2\phi-\phi_S)}$ becomes 
\begin{linenomath}
\beq
\begin{split}
A_{UT,L}^{\sin(2\phi-\phi_S)} & =-\frac{\sqrt{\epsilon(1+\epsilon)}}{r^{04}_{00}}\,\rm{Im}[n^{00}_{0+}+s^{00}_{0+}]\\
& =-\frac{\sqrt{\epsilon(1+\epsilon)}}{u^{00}_{++}+\epsilon u^{00}_{00}}\,\rm{Im}[n^{00}_{0+}+s^{00}_{0+}].
\label{def-a6l}
\end{split}
\eeq
\end{linenomath}
Correspondingly, for the production of transversely polarized vector mesons, the terms with $\mu_1=\mu_2=\pm 1$ have to 
be retained in Eqs.~(\ref{a1_o}--\ref{sum-s}), and the result has to be divided by  
$(1-r^{04}_{00})$. For instance, $A_{UT}^{\sin(2\phi-\phi_S)}$ becomes 
\begin{linenomath}
\beq
\begin{split}
& A_{UT,T}^{\sin(2\phi-\phi_S)} \\
& \quad =-\frac{\sqrt{\epsilon(1+\epsilon)}}{1-r^{04}_{00}}\,\rm{Im}[n^{++}_{0+}+s^{++}_{0+}+n^{--}_{0+}+s^{--}_{0+}]\\
& \quad =-\frac{\sqrt{\epsilon(1+\epsilon)}}{1-u^{00}_{++}-\epsilon u^{00}_{00}}\,\rm{Im}[n^{++}_{0+}+s^{++}_{0+}+n^{--}_{0+}+s^{--}_{0+}].
\label{def-a6t}
\end{split}
\eeq
\end{linenomath}
\end {appendix}

\end{document}